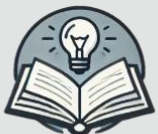

# The "strength" of patent systems

Gaétan de Rassenfosse

Holder of the Chair of Science, Technology, and Innovation Policy
École polytechnique fédérale de Lausanne, Switzerland.

This version: May 2025

**Purpose**
This article is part of a Living Literature Review exploring topics related to intellectual property, focusing on insights from the economic literature. Our aim is to provide a clear and non-technical introduction to patent rights, making them accessible to graduate students, legal scholars and practitioners, policymakers, and anyone curious about the subject.

**Funding**
This project is made possible through a Living Literature Review grant generously provided by Open Philanthropy. Open Philanthropy does not exert editorial control over this work, and the views expressed here do not necessarily reflect those of Open Philanthropy.

# The "strength" of patent systems

Gaétan de Rassenfosse
École polytechnique fédérale de Lausanne, Switzerland.

As discussed in a previous article, patent "quality" has many facets.[1] One is compliance with statutory requirements. Do patent offices grant and refuse applications in line with their own rules, or do they make mistakes? Under this lens, a "low-quality" patent is one unlikely to survive a court challenge. But perfect compliance is not the whole story. A patent system could have a bar so low that the office issues patents on every trivial invention. The office would never make a "mistake," yet such a low threshold is hardly a mark of true quality.

That discussion brings us to a second, equally important dimension: rigor. How high is the bar? Are patents awarded only for genuine advances—novel, non-obvious contributions—or is the threshold so low that almost anything under the sun qualifies as "inventive"? To capture these differences, scholars have devised indices measuring the strength of patent systems. This article reviews some of the major indices and explores what they reveal about international variation in patent systems.

**Ginarte and Park's index**

The Ginarte and Park index was first released in 1997 and remains the most widely used measure of patent-system strength. Drawing on the formal laws "on the books," it assigns each country a composite score based on five legislative and institutional dimensions:

1. *Coverage*. Assesses which categories of inventions are eligible for patent protection (*e.g.*, pharmaceuticals, chemicals, plant varieties, software, *etc.*). Fewer exclusions imply a more expansive ("stronger") system.
2. *Membership in international treaties*. Awards points for participation in key agreements such as the Paris Convention, the Patent Cooperation Treaty, and the International Convention for the Protection of New Varieties of Plants. Treaty obligations typically raise minimum standards and promote harmonization.
3. *Duration of protection*. Measures the statutory patent term. The international norm of 20 years from filing serves as the benchmark; shorter terms reduce a country's score.
4. *Enforcement mechanisms*. Captures whether national law provides tools for patent holders to defend their rights, such as preliminary injunctions, contributory-infringement claims, reversals of the burden of proof, and similar remedies.
5. *Restrictions on patent rights*. Penalizes systems that impose compulsory licensing, domestic-working requirements, or other legal limits that weaken patent exclusivity.

By summing the scores across these five categories, the Ginarte and Park index delivers a single number reflecting each country's formal patent-law strength. Higher scores indicate more robust regimes—that is, systems with broader coverage, stronger enforcement, and fewer built-in limitations. This index is widely employed in economic research to examine how variations in patent-law strength affect indicators such as innovation, trade flows, and foreign direct investment (FDI).

---

[1] de Rassenfosse, G. (2025). What is patent quality? The Patentist Living Literature Review 4: 1–5. DOI: 10.48550/arXiv.2504.08785.

Yet, despite its broad adoption, scholars have identified notable limitations of the Ginarte and Park index. First, as a general measure, it may obscure sector-specific features of patent regimes. This issue prompted the creation of targeted indices, such as those for pharmaceuticals (Liu and La Croix 2015) and plant varieties (Campi and Nuvolari 2015). Second, by relying solely on "law on the books," it overlooks the real-world effectiveness of those laws—what scholars call "law in practice." Third, because it focuses on the generosity of formal rules and remedies, calling high scores "strong" can be misleading: these values often reflect applicant friendliness more than patent system robustness. In the sections that follow, we explore the challenges of measuring enforcement in practice and consider alternative approaches to assessing patent quality.

**Law in practice**

The Agreement on Trade-Related Aspects of Intellectual Property Rights (TRIPS), administered by the World Trade Organization (WTO), entered into force on January 1, 1995. It is the most comprehensive multilateral treaty on intellectual property (IP), setting minimum standards for the availability, scope, and use of IP rights. TRIPS spurred rapid convergence in "law on the books" across WTO members, but significant disparities remain in enforcement and administration. Patent owners' real ability to exercise their rights depends not only on statutory rules but on how those rules are applied in practice. Hence, the need for a complementary perspective: law in practice, which captures the effectiveness of IP-law enforcement by courts, police, customs authorities, and other institutions.

Because "law in practice" involves diffuse enforcement routines, nuanced procedural norms, and informal institutional behaviors that resist simple codification, it is far harder to measure than "law on the books." Direct data on enforcement outcomes—such as court filings, injunctions granted, or damages awarded—are often unavailable or incomplete. Even where records exist, compiling comparable indicators of judicial independence, administrative efficiency, or enforcement-official competence across dozens of countries is no small feat. Layered on top of this are culture-bound conventions and unwritten protocols—how judges, prosecutors, and customs officers actually interpret and apply statutes day to day—which defy straightforward quantification.

Despite these challenges, researchers have devised several proxies for "law in practice." Ostergard (2000), for example, incorporated enforcement evidence drawn from the U.S. State Department's Country Reports on Economic and Trade Practices. Javorcik (2004) used assessments from the International Intellectual Property Alliance's Special 301 recommendations. Papageorgiadis et al. (2014) offer perhaps the most comprehensive analysis, with an index that quantifies enforcement based on three categories of transaction costs faced by rights holders:

1. *Servicing costs.* Quality of patent administration.
2. *Property-rights protection costs.* Judicial efficiency, corruption, and related factors that affect court-based enforcement.

3. *Monitoring costs.* Resources required to detect infringement, including police engagement and public commitment to IP protection.

**"Strength," or applicant-friendliness?**

The indices we have reviewed so far capture the strength of patent laws and enforcement, but they don't tell us whether the patents actually granted would survive a court challenge. In other words, they measure how *generous* or *friendly* a system is to patent owners, not how *robust* its patents are.

To address this gap, de Saint-Georges and van Pottelsberghe (2013) developed a "quality index" built around nine operational design features. Seven of these relate to procedural safeguards—such as the window for requesting examination, the availability of post-grant opposition, and grace-period length—while two indicators capture patent office resourcing: personnel expenses per staff member and the average number of claims examined per examiner.

Building on this procedural focus, Gimeno-Fabra and van Pottelsberghe (2021) convert procedural information published by patent offices into quantifiable examiner metrics, including search completeness through classification and citation practices, the upfront certainty of information delivery, the speed of information delivery, and the relative stringency assessed via grant rates. This research line admittedly offers a more direct gauge of patent strength. However, it does not tell us how high or low the bar is for patent applications to be granted.

**Insights from "twin patent" studies**

A complementary approach to gauging patent-system rigor uses twin patents—matched filings for the same invention made at different offices—and compares grant outcomes across jurisdictions. Jensen et al. (2006), for example, examine triadic patents (filed in the United States, Japan, and Europe) and document significant "disharmony": decisions on the same invention often diverge, with some offices granting patents that others reject.

Such inconsistencies are not inherently problematic, for different countries may legitimately apply distinct patentability standards. However, they can also signal examiner errors. In a follow-up study, de Rassenfosse et al. (2021) decompose these discrepancies into two sources: variation in patentability thresholds and outright mistakes. Analyzing data from the five largest offices (the triadic offices plus China and South Korea), they find that systemic differences in examination standards explain most disharmony, while true errors play a more limited role.

Crucially, their model also quantifies each office's effective "bar height." They show that the JPO applies the strictest standard, followed by the EPO, the KIPO, the USPTO, and finally CNIPA—clearly documenting differences in patent offices' rigor.

**What explains cross-country differences in patent protection?**

Recognizing the vast differences in the level of patent protection across countries, scholars have sought to understand the factors that explain these variations. They find that countries tend to calibrate their IP regimes dynamically as they develop. At early stages of industrialization, when domestic firms lack significant innovative capacity, patent systems are often weak or narrow. Weak systems allow local actors to imitate established technologies without stringent legal barriers. As a nation's technological base and R&D capabilities grow, however, pressure mounts to strengthen IP rights, both to reward nascent innovators and to secure access to foreign markets. China exemplifies this trajectory: its patent laws and enforcement apparatus have been progressively tightened over the last three decades, in tandem with the rise of home-grown "national champions" that compete globally and demand robust protection at home.

However, whether weak IP regimes truly benefit developing economies remains a topic of hot debate. Critics of lax IP argue that multinational firms will hesitate to establish R&D centers or high-value manufacturing in countries where their core technologies can be pirated. Empirical studies have found that stronger patent protection is correlated with greater inflows of technology-intensive FDI, suggesting that weak regimes can deter knowledge transfer and thereby limit learning-by-doing opportunities.

However, proponents of early-stage flexibility counter that imitation is a key engine of catch-up. By allowing domestic firms to reverse-engineer and adapt foreign innovations, weak IP rights can accelerate capability building, foster competitive industries, and lower entry barriers.

Ultimately, the political economy of IP reform—shaped by domestic interest groups, trade negotiations, and institutional capacity—determines each country's unique path along this dynamic spectrum.

## References

Campi, M., & Nuvolari, A. (2015). Intellectual property protection in plant varieties: A worldwide index (1961–2011). Research Policy, 44(4), 951–964.

de Rassenfosse, G., Griffiths, W. E., Jaffe, A. B., & Webster, E. (2021). Low-quality patents in the eye of the beholder: Evidence from multiple examiners. The Journal of Law, Economics, and Organization, 37(3), 607–636.

de Saint-Georges, M., & van Pottelsberghe de la Potterie, B. (2013). A quality index for patent systems. Research Policy, 42(3), 704–719.

Gimeno-Fabra, L., & van Pottelsberghe de la Potterie, B. (2021). Decoding patent examination services. Economics of Innovation and New Technology, 30(7), 707–730.

Ginarte, J. C., & Park, W. G. (1997). Determinants of patent rights: A cross-national study. Research Policy, 26(3), 283–301.

Javorcik, B. S. (2004). The composition of foreign direct investment and protection of intellectual property rights: Evidence from transition economies. European Economic Review, 48(1), 39–62.

Jensen, P. H., Palangkaraya, A., & Webster, E. (2006). Disharmony in international patent office decisions. Federal Circuit Bar Journal, 15, 679–704.

Liu, M., & La Croix, S. (2015). A cross-country index of intellectual property rights in pharmaceutical inventions. Research Policy, 44(1), 206–216.

Ostergard, R. L. (2000). The measurement of intellectual property rights protection. Journal of International Business Studies, 31(2), 349–360.

Papageorgiadis, N., Cross, A. R., & Alexiou, C. (2014). International patent systems strength 1998–2011. Journal of World Business, 49(4), 586–597.